\documentclass[prl,a4paper,aps,twocolumn,superscriptaddress]{revtex4-1}

\usepackage[utf8]{inputenc}
\usepackage[english]{babel}
\usepackage[T1]{fontenc}
\usepackage{amsmath,times}
\usepackage{amsfonts}
\usepackage{ulem,soul}

\usepackage[pdftex,breaklinks,colorlinks=true,allcolors=blue]{hyperref}

\usepackage {subfigure}
\usepackage{color}
\usepackage {overpic}
\usepackage{xcolor}

\definecolor{commentgreen}{RGB}{2,112,10}

\usepackage{listings}
\def\bra#1{\mathinner{\langle{#1}|}}
\def\ket#1{\mathinner{|{#1}\rangle}}

\def\bra#1{\mathinner{\langle{#1}|}}
\def\ket#1{\mathinner{|{#1}\rangle}}

\begin{document}
\title{Continuous-time quantum walks on spatially correlated noisy lattices}
\date{\today}
\author{Matteo A. C. Rossi}
\affiliation{Quantum Technology Lab, Dipartimento di Fisica, Universit\`a degli Studi di Milano, I-20133 Milano, Italy}
\author{Claudia Benedetti}
\affiliation{Quantum Technology Lab, Dipartimento di Fisica, Universit\`a degli Studi di Milano, I-20133 Milano, Italy}
\author{Massimo Borrelli}
\affiliation{Turku Centre for Quantum Physics, Department of Physics and Astronomy, University of Turku, FI-20014 Turun yliopisto, Finland}
\author{Sabrina Maniscalco}
\affiliation{Turku Centre for Quantum Physics, Department of Physics and Astronomy, University of Turku, FI-20014 Turun yliopisto, Finland}
\affiliation{Center for Quantum Engineering, Department of Applied Physics, Aalto University School of Science, FI-00076 Aalto, Finland}
\author{Matteo G. A. Paris}
\affiliation{Quantum Technology Lab, Dipartimento di Fisica, Universit\`a degli Studi di Milano, I-20133 Milano, Italy}
\begin{abstract}
We address memory effects and diffusive properties of a continuous-time quantum
walk on a one-dimensional  percolation lattice affected by spatially correlated
random telegraph noise. In particular, by introducing spatially correlated
time-dependent fluctuations in nearest-neighbor hopping amplitudes, we describe
random domains characterized by global noise. The resulting open dynamics of
the walker is then unraveled by an ensemble average over all the noise
realizations. Our results show that time-dependent noise assisted by
spatial correlations leads to strong memory effects in the walker
dynamics and to robust diffusive behavior against the detrimental
action of uncorrelated noise. We also show that spatially correlated classical
noise enhances localization breaking, thus making a quantum particle
spread on longer distances across the lattice.
\end{abstract}
\maketitle
{\it Introduction -} Continuous-time quantum walks (CTQWs) describe the free
evolution of quantum particles on $N$-vertex graphs. They have been subject
of intense studies, both theoretical and experimental, as they have proven
useful for several applications, ranging from universal quantum computation \cite{childs09}, to search algorithms \cite{childs04, omar16},  quantum transport \cite{mulken07,Bougroura16}, quantum state transfer \cite{tama16} and energy
transport in biological systems \cite{Mohseni08}.
\par
Given their relevance in applications, a realistic description of the dynamics
of quantum walkers should take into account those sources of noise and imperfections that might jeopardize the discrete lattice on which the CTQW occurs. While the effects of both disorder and dynamical fluctuations have been analyzed in the recent past \cite{yin08,schreib11,rwclaudia,siloi17,li15,Chattaraj16}, the consequences
of spatially-correlated noise on the dynamics of the walker are still,
to the best of our knowledge, an unexplored territory.
\par
In this paper, we address the effects  of spatially-correlated noise by studying
the most relevant dynamical features of a one-dimensional CTQW affected both by
time- and space-dependent fluctuations. As for the former, the hopping amplitudes
are assumed to fluctuate in time as a random telegraph noise (RTN) inducing dynamical percolation, which results
in a stochastic time-dependent Hamiltonian. This model has been studied in \cite{rwclaudia} in connection to particle localization and memory effects
in the open dynamics of the walker. Here, we take a step further and introduce
random spatial correlations as follows: if two adjacent hopping links are
subject to spatially correlated fluctuations, then they are affected by the
same RTN time evolution.
This
will lead to the formation of percolation domains within which the tunneling amplitudes
evolve according to the same stochastic noise. On a global scale, because
of these  spatial correlations, the hopping fluctuations will be synchronized
domain-wise. Overall, this is perhaps the simplest type of space dependency that
one may introduce to the 1D CTQW, as it does not interfere with the local time-dependent part of the noise. The two sources of noise correlations may indeed
be treated independently. At the same time, the model allows one to describe
the formation of spatial domains and to address percolation effect.
\par
In turn, the motivation for introducing this extra ingredient is two-fold. First,
if one aims at a more realistic description of any experimental implementation of
a CTQW, sources of noise should be accounted for.  This is especially important
when studying transport properties in disordered systems in which
localization, let it be Anderson or many-body \cite{anderson,manybloc,manyblocrev}, represents an obvious obstacle. A renewed interest in this field has spurred deep investigations
in highly-engineered experimental setups, such as cold atoms in optical
lattices \cite{blochrev,cazalilla}, in which complex noise might be efficiently implemented \cite{floquet}.
The second aspect concerns the question of whether the introduction of
spatially correlated noise might result in improving certain dynamical
features, such as slowing down decoherence or even enhancement
of quantum properties. In this respect, memory effects are of primary
importance, as they have been shown to improve the performances of numerous
protocols in quantum information \cite{antti,biheng,laine2014,bogna2} and quantum
metrology \cite{metro1,metro2}. They also play a key role in
quantum thermodynamics \cite{bogna} and measurement theory
\cite {karpat2015}. However, non-Markovian dynamics has been so far
 widely investigated and understood by focusing on the time/frequency domain \cite{reviewopen}, e.g. by
inspecting quantities such as correlation functions at different times and
spectral densities of certain environments. It is this not obvious how, and
whether, introducing spatially dependent noise might affect memory effects
of a given dynamical map.
\par
Our findings shed light on the effect of space-correlated dynamical
noise on a quantum map, in terms of both memory effects and kinetic quantities,
such as diffusion and velocity. As a matter of fact, spatially correlated noise
results in stronger memory effects in the dynamics and it partly suppresses the
localization induced by its randomness, allowing the walker to spread further
and faster across the lattice.
\par
{\it The model -} We consider a 1D lattice of $N$ sites and one particle
(walker) freely moving across it. The Hamiltonian $H$ describing the walker's
dynamics can be expanded in the single-particle localized orthonormal basis
$\{\ket{j}\}$ with $j = 1,\dots, N$. If we introduce
time-dependent stochastic fluctuations on the hopping amplitudes, the
time-dependent Hamiltonian $H(t)$ reads:
\begin{equation}
H(t)=-\sum_{j}\big[\nu_{0}+\nu g_{j}(t)\big]\Big(|{j}\rangle\!\langle{j+1}|+|{j+1}\rangle\!\langle{j}| \Big),
\label{hamiltonian}
\end{equation}
in which  $\nu_0$ is the uniform hopping amplitude between nearest neighbor sites,
$\nu$ is the noise strength and $\{g_{j}(t)\}_j$ are independent RTN processes
that jump between $\pm 1$ according to the switching rate $\gamma$.
\par
\begin{figure}[t]
\centering
\includegraphics[width=1.\columnwidth]{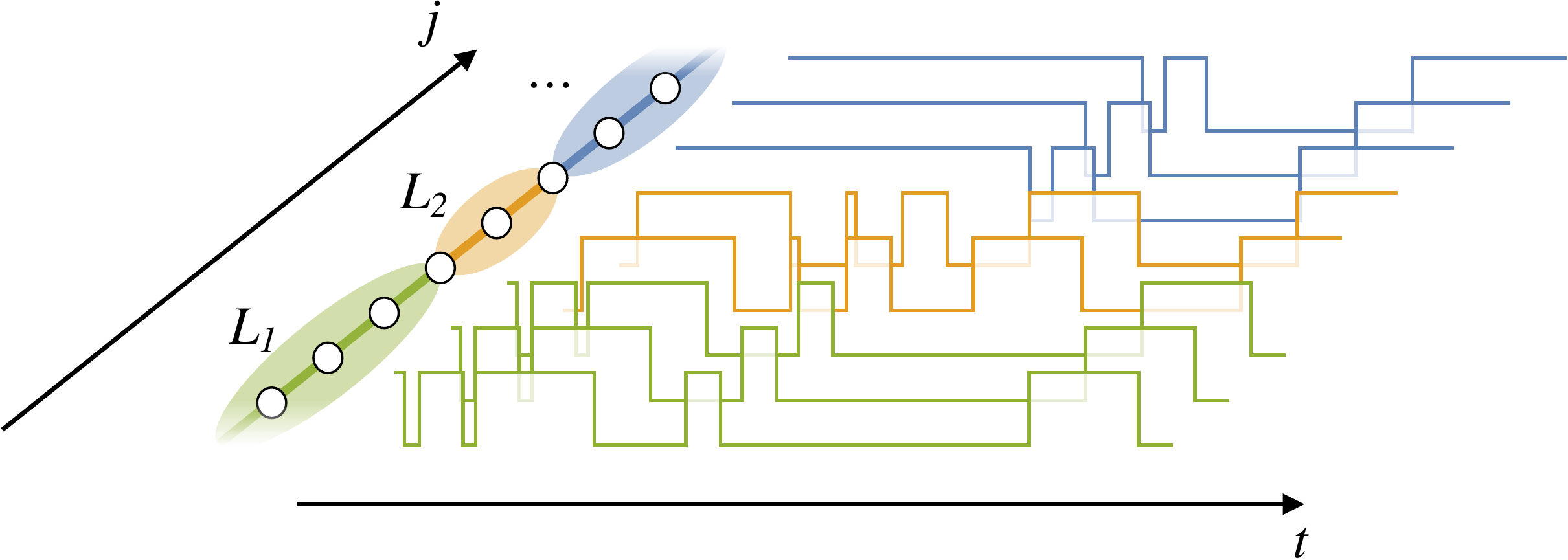}
\caption{Schematic representation of the random spatial domains $\{L_1, L_2,\dots, L_M\}$ for a single realization of the noise, generated according to Eq. \eqref{binomial} and of average length ${\bar L}_p$. Tunneling amplitudes within the same domain fluctuate synchronously in time and according to the same stochastic process. Different domains evolve independently from each others. }
\label{schema}
\end{figure}

We now introduce  spatial correlations in the noisy Hamiltonian \eqref{hamiltonian} as follows. We assume that two adjacent links of the lattice can be noise-correlated
with a certain probability $p$. Formally, this translates to the following autocorrelation function:
\begin{equation}
    \langle g_{j}(t)g_{k}(0)\rangle=
\begin{cases}
    \propto e^{-2\gamma t},& \text{if } j,k \;\textrm{correlated}\\
    0,              & \text{otherwise}
\end{cases}
\label{CorrFunc}.
\end{equation}
For a single noise realization, these spatial correlations
will form $M$ domains  of lengths $\{L_{1},L_{2},\dots,L_{M}\}$, corresponding to $M$ independent noise evolutions $\{g_{1}(1),g_{2}(t),\dots,g_{M}(t)\}$ respectively, as shown in Fig. \ref{schema}.
The distribution of the domains is random and different for each noise realization: the probability $P_{M}$ of having $M$ domains  in a particular noise realization is described by a binomial distribution
\begin{equation}
P_{M}=\binom{N-1}{M-1}(1-p)^{M-1}p^{N-M},
\label{binomial}
\end{equation}
which corresponds to the following average domain length $\bar{L}$
(as a function of $p$) $\bar{L}_{p}=\frac{p^{N}-1}{p-1}$.
By continuity, we define $\bar{L}_{1} = \lim _{p\to1} \bar{L}_{p}=N$. In this case, there is a single noise domain that spans the whole lattice.

So far, the amplitude of the fluctuations $\nu$ has been considered a free parameter of the strength of the noise. Here, we are interested in the effect of noise space  and time correlations {\it per se}, rather than in the noise strength. Thus, we set this parameter to $\nu=\nu_0$, meaning that, from now on, we are only going to consider { percolation} noise: the local hopping amplitudes can switch between 0 and $2\nu_{0}$ \cite{Darazs13}, resulting in links that are created and destroyed randomly in time, according to the statistics of the RTN process.
Quite obviously, this analysis can be carried out for any value of $\nu$.
\par
For each  noise realization, the system time evolution is ruled by the operator $U(t)=\mathcal{T}e^{-i\int_{0}^{\tau}d\tau H(\tau)}$. The open dynamics of the walker is unraveled by computing the ensemble average of the unitary dynamics over all possible realizations
\begin{equation}
\bar{\rho}(t)=\Lambda_t\,\rho_0=\langle U(t)\rho_0U^{\dagger}(t) \rangle_{\{g(t)\}}
\label{averagestate}
\end{equation}
where $\langle . \rangle_{\{g(t)\}}$ indicates the average taken over an (in principle) infinite number of implementations of the sets $\{g_{1}(1),g_{2}(t),\dots,g_{M}(t)\}$ and $\rho_0$ is the (fixed) initial state of the walker.
Needless to say, whenever the solution to Eq.~\eqref{averagestate} is analytically out of reach, one can only numerically approximate this ensemble-average with a finite number of noise realizations $R$. In this case we talk about  under-sampling \cite{cialdi17} and the true dynamics \eqref{averagestate} can be recovered only in the limit $R\rightarrow\infty$.
For all the quantities computed in this work the size of the noise sample is $ R=10\,000$, which guarantees statistical robustness of our results. The code for simulating the dynamics is reported and explained in the Appendix.
\par {\it Non-Markovianity of the dynamical map -} As previously mentioned, the noise-averaged dynamics of the walker can no longer be described by Schr\"{o}dinger equation
and one has to resort to the machinery of open quantum systems. In this respect, a relevant question is whether the
open dynamics of the walker is memory-less, i.e. Markovian, or  non-Markovian.
In Ref.~\cite{rwclaudia} memory effects in the dynamics
of the walker in presence of spatially uncorrelated RTN were investigated for some selected initial states
leading to the conclusion that decreasing the switching rate $\gamma$ enhances the memory effects.
That scenario corresponds to noise domains of average length $\bar{L}=1$ and therefore it is a special
case study of the more general  model introduced in this manuscript.
Intuitively, since the non-Markovian dynamics is intrinsically connected to the time-dependency of the environment correlation functions, we can expect that whenever the spatial-uncorrelated noise is Markovian, it will also be Markovian in the spatially-correlated-noise case. This is simply because, as mentioned previously, the spatial correlations in the noise
do not interfere with the RTN itself but they only {\it assist} it.
However, if memory effects are  present already in the spatially uncorrelated scenario, it is not
obvious {\it a priori} how long-range correlated  noise with $\bar{L}>1$ will affect the non-Markovianity of the quantum map.
Similarly to \cite{rwclaudia}, we use the trace-distance-based Breuer-Laine-Piilo (BLP) \cite{blp} approach
to characterize memory effects in the open dynamics of the walker. The trace distance between two quantum states $\rho_{1}$ and $\rho_{2}$ is defined as
$
D(t)=D\big(\rho_{1}(t),\rho_{2}(t)\big)=\frac{1}{2}||\rho_{1}(t)-\rho_{2}(t)||
$, where
$||A||=\textrm{Tr}\left[\sqrt{A^{\dagger}A}\right]$, $\rho_{1(2)}(t)=\Lambda_t \rho_{1(2)}$ and $\Lambda_t$ denotes
a dynamical quantum map. For a Markovian map, $D(t)$  monotonically decreases in time for any initial pair of states.
Therefore, a violation of such a constraint signals the presence of memory-effects or, equivalently, a
non-Markovian dynamics.
A quantifier of memory effects can be defined by integrating the time derivative of $D(t)$ over the time intervals
where the trace distance has revivals, i.e.  $\dot{D}(t)=dD(t)/dt > 0$, and then maximizing over all the possible pairs of initial states.
Computationally, this translates to evaluating the following quantity:
\begin{equation}
\mathcal{N}=\textrm{max}_{\rho_{1},\rho_{2}}\int_{\dot{D}(t)>0} \!\!\!
dt\, \frac{d}{dt}D(\Lambda_{t}\rho_{1},\Lambda_{t}\rho_{2})
\label{nblp}
\end{equation}
in which $\Lambda_{t}$ is the dynamical map \eqref{averagestate}. The above
quantity is, in practice, nearly impossible to compute exactly because it involves a state optimization procedure and only few
analytically treatable cases are known in literature \cite{antti}. Nonetheless, it does provide a rather intuitive interpretation of memory effects in
open systems and it still allows to get an insight of the behavior of memory effects by selecting some significant pairs of initial states.
\par {\it Diffusion  vs localization: the inverse participation ratio -} In Ref. \cite{rwclaudia}, the dynamics governed by Eq.~\eqref{hamiltonian} in absence
of spatially correlated noise was analyzed in details, showing  a transition from a diffusive to a localized
regime as a function of the switching rate  $\gamma$. Furthermore, depending
on the strength of the noise, a quantum-to-classical transition was also observed for the fast noise
case ($\gamma>1$), resulting in a  Gaussian probability distribution of the walker's state.
Here, we aim at understanding the role of noise spatial correlations in the dynamical behavior of the walker. Specifically, we want to understand whether spatially correlated noise domains help the particle spread over the lattice or whether they instead favor localization.
We quantify the extent of noise-induced localization by means of the  inverse participation
ratio (IPR) \cite{ipr}, defined as
\begin{equation}
\mathcal{I}(t)=\sum_{j=1}^{N}\bra{j}\bar{\rho}(t)\ket{j}^{2}.
\label{ipr}
\end{equation}
IPR is bounded between $1/N$ and 1 with $\mathcal{I}(t)=\frac{1}{N}$ meaning
complete delocalisation and $\mathcal{I}(t)=1$ corresponding to
localization on a single site. The larger the IPR, the more localized the
particle is. Using IPR we now investigate how the spatially correlated time-dependent
noise affects the diffusive properties of the walker.
\begin{figure}[h!]
  \includegraphics{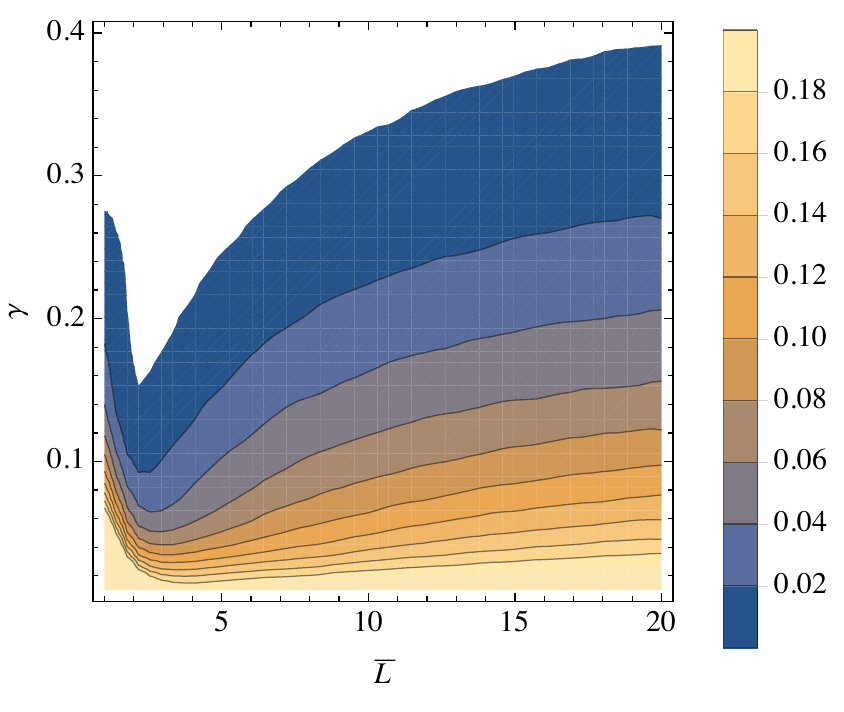}
  \caption{Non-Markovianity $n_{\tau}\left(\gamma,\bar{L}\right)$ as a function of the
  average domain length $\bar{L}$ and switching rate $\gamma$ for percolation noise.
  The selected initial states are $\ket{N/2}$ and $\ket{N/2+1}$ with $N=100$ and $\nu_0\tau= 20$.
  In the white region, $n_{\tau}\left(\gamma,\bar{L}\right) = 0$.}
  \label{PlotNonMark}
\end{figure}
\par {\it Results} - We now present our results on the dynamical properties
of the walker in a noisy, spatially correlated lattice. The evolution of the
walker is obtained by randomly generating the domains and the
noise realizations, computing the single realization unitary dynamics and finally performing the ensemble average \eqref{averagestate} for a $N=100$ lattice and for $R=10\,000$ iterations.
First, we will focus on the non-Markovian character of the quantum map, then analyze the diffusive properties of the CTQW.
As anticipated above, the maximization in Eq.~\eqref{nblp} is
a nearly impossible task for most physical systems. Because of our
computational resources and the complexity of the model at hand,
this case-study is certainly no exception.
However, we can still compute the integral in Eq.~\eqref{nblp} for some relevant
initial pairs of states and gain useful information regarding at least their dynamics.
Since we are interested in the interplay between noise-induced localization and memory
effects due to spatially correlated noise, we restrict our attention to pairs
of initial states that are localized on adjacent sites and we compute the
following quantity
\begin{equation}
n_{\tau}\left(\gamma,\bar{L}\right)=\int_{\!\dot{D}(t)>0}\!\!\!  dt \,\frac{d}{dt}D\big(\Lambda_{t}\rho_{N/2},\Lambda_t\rho_{1+N/2}\big),
\label{blpLoc}
\end{equation}
for a fixed final time $\tau$, as a function of $\gamma$. In the above equation, $\rho_{j}=\ket{j}\!\!\bra{j}$ and $\Lambda_t=\Lambda_t{(\gamma,\bar{L})}$ is the dynamical map computed via Eq.~\eqref{averagestate} that depends upon the value of the noise switching rate $\gamma$ and the average domains length $\bar{L}$.
The integral over time in Eq. \eqref{blpLoc} is up to the fixed time $\tau$.
In Fig.~\ref{PlotNonMark} we display $n_{\tau}\left(\gamma,\bar{L}\right)$ for a $N=100$ lattice and $\nu_{0}\tau=20$. We choose this truncation time to ensure that the tails of the walker wave-function
have not yet reached the boundaries of the lattice and therefore we need not to worry about finite-size-induced memory effects. Here we analyze a range of values for $\gamma$ that are known to generate non-Markovian dynamics, for the same initial states, in the case of non-correlated RTN \cite{rwclaudia}.
The striking feature we immediately notice is that, after a minimum located at $\bar{L}\approx 2$ and independent of $\gamma$, as the average domain length $\bar{L}$ is increased, the non-Markovian character evaluated through Eq. \eqref{blpLoc} also increases. Thus spatial correlations in the noise make memory effects stronger, at least for this set of initial states.
An intuitive theoretical explanation of this behavior might be the following. The presence of domains with a typical length $\bar L$ is effectively equivalent to amplifying the single-link contribution to memory effects proportionally to the size of the domain. The walker experiences a smaller effective lattice of size $M$ with, however, stronger average local disorder. We performed this calculation using increasingly separated localized initial states and found the exact same behavior, with the only difference being a smaller value of $n_{\tau}\left(\gamma,\bar{L}\right)$.
\par
\begin{figure}[t]
\centering
  \includegraphics{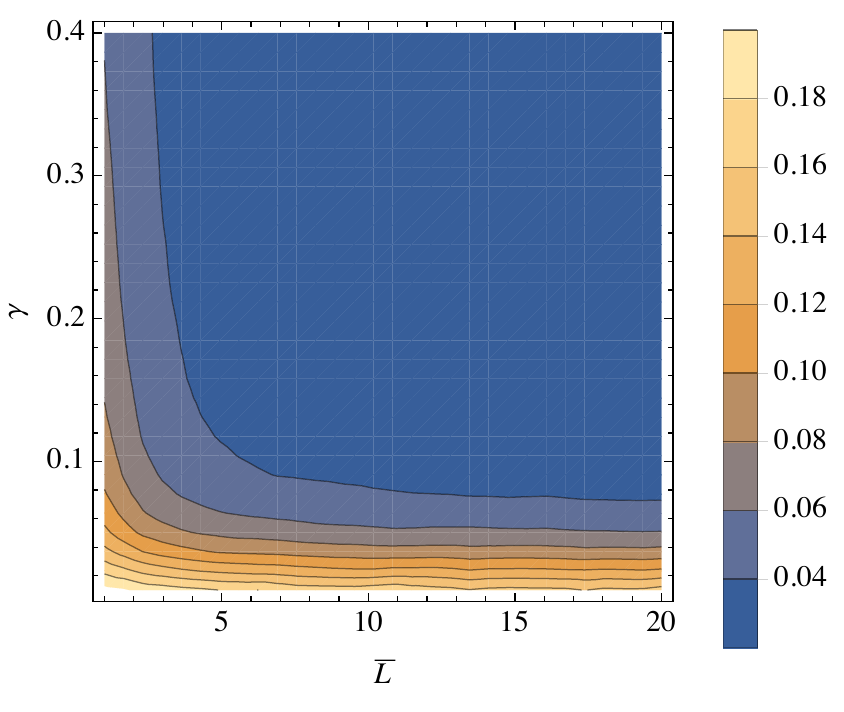}
  \caption{Long-time value of the IPR as a function of average domain length $\bar{L}$ and switching rate $\gamma$ for percolation noise for the initial states $\ket{N/2}$  with $N=100$ for $\nu_0\tau= 20$.}
  \label{PlotIPRLoc}
\end{figure}

In Fig.~\ref{PlotIPRLoc}, we display the long-time IPR $\mathcal{I}(\nu_0\tau)$ as a function of $\gamma$ and $\bar{L}$ computed for the initially localized state $\ket{N/2}$.
This quantity is defined as the quasi-stationary value that the IPR reaches before boundary effects come into play. Interestingly, for a fixed $\gamma$, this has a maximum at $\bar{L}=1$, similarly to $n_{\tau}\left(\gamma,\bar{L}\right)$, and decays fast as $\bar{L}$ increases. While uncorrelated slow noise tends to keep the walker localized around its initial position, spatial correlations break the localization and lead to  a stronger diffusion of the wave function across the lattice.  By increasing the value of the switching rate $\gamma$, the IPR  becomes smaller as we approach
memory-less and more diffusive dynamics. Therefore, while the presence of slow noise  ($\gamma<1$) tends to favor localization, by adding random spatial correlations to the
very same noise we can limit this effect and allow the walker to propagate through the lattice while still retaining memory effects in its dynamics. Overall, and perhaps quite unexpectedly, for a small fixed $\gamma$, a spatially-correlated RTN tends to suppress localization while still enhancing memory effects.
\par
To investigate transport properties in this setting we turn our attention to an initial Gaussian wave-packet, equipped with an average momentum $k_{0}$ and spatial spread $\Delta$
\begin{equation}
\ket{\mathcal{G}}=\sum_{j=1}^{N}\left[\frac{1}{\sqrt{2\pi\Delta^{2}}}e^{-\frac{\left(j-\frac{N}{2}\right)^{2}}{2\Delta^{2}}}\right]e^{-ik_{0}j}\ket{j}.
\label{gaussian}
\end{equation}
\begin{figure*}
\centering
\includegraphics{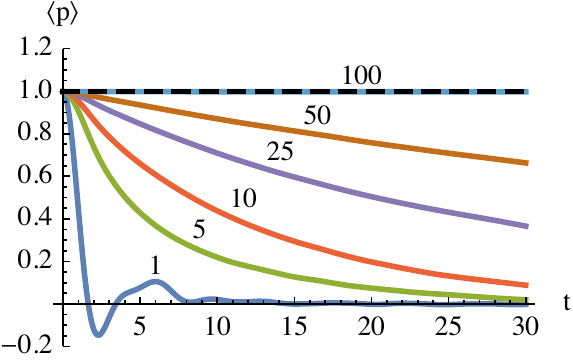}
\includegraphics{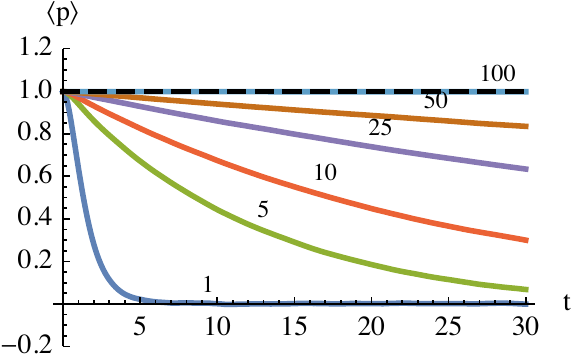}
\includegraphics{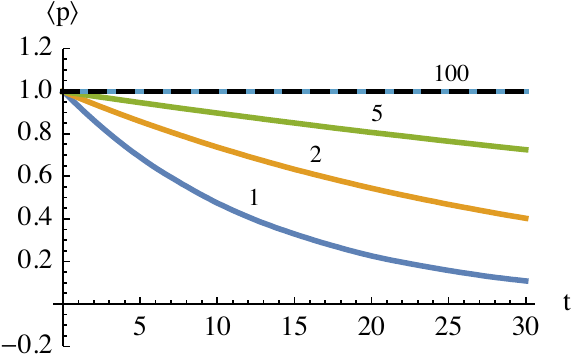}
\vskip 1em
\includegraphics{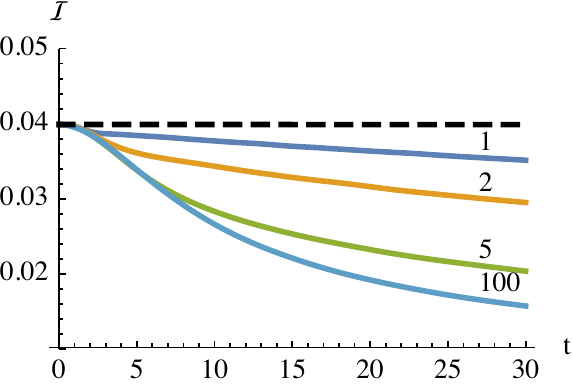}
\includegraphics{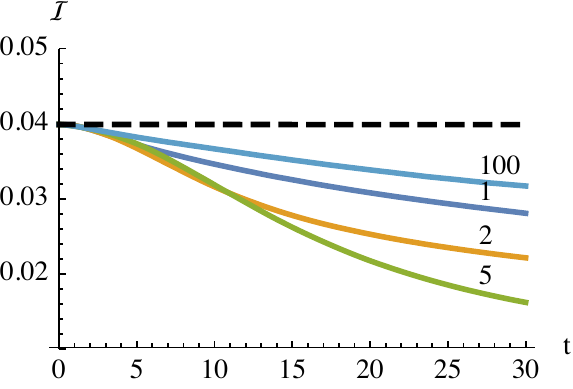}
\includegraphics{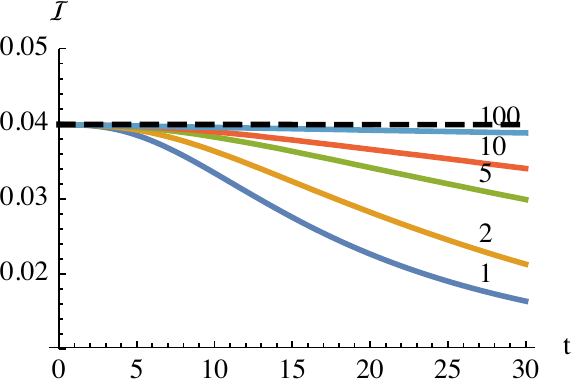}
\caption{ Expectation value of the momentum operator $\langle p \rangle$ (top panels) and IPR $\mathcal{I}$ (bottom panels) as a function of time,
for different average domain lengths $\bar{L}$, for $\gamma=0.1$ (left), 1 (center) and 10 (right), with lattice size $N=100$. The black dashed line indicates the noiseless case. The initial state is \eqref{gaussian}, with $k_0= \pi/2$, $\Delta=10$.}
\label{paverage}
\end{figure*}
 We  study the behavior of both the IPR and the average momentum operator $\hat{p}=-i \nabla$, computed using the Born rule $\langle \hat{p}(t) \rangle= \text{Tr}\big[\bar{\rho}(t)\,\hat{p}\big]$, which represents the average quantum velocity at which the wave packet travels across the lattice. Figure \ref{paverage} shows the time evolution of these two quantities for three different values of the switching rate $\gamma$ and different average domain lengths $\bar{L}$.
 In this case, the effects of the spatially correlated RTN become even clearer.
The wave-packet momentum $\langle p \rangle$ (upper panel) decreases in time, until it
eventually vanishes asymptotically, and this decay is faster for smaller values of $\gamma$, in agreement with Fig.~\ref{PlotIPRLoc}. However, while space-uncorrelated noise leads to a faster reduction of $\langle \hat{p}\rangle$, spatial correlations in the RTN allow  the wave-packet to preserve momentum and travel longer across the lattice before stopping. In the limiting case of $\bar L = N$ (i.e. $p=1$), the average momentum $\langle \hat{p}\rangle$ is preserved, as in the noiseless case.

Similarly to the case studied above, the IPR (lower panel) generally decreases in time.
However, there seems to exist a more complicated interplay between $\gamma$ and $\bar{L}$.
For small $\gamma$ the IPR decays faster for larger values of $\bar{L}$, indicating that spatial correlations
break the noise-induced localization, in agreement with our previous results.
For larger switching rates $\gamma$, instead, the situation is quite the opposite: strong spatial correlations
prevent the particle distribution from spreading further, thus preserving the initial IPR, with the limiting case of $p=1$, i.e. $\bar L = N$ that gives the slowest possible decay.

Since the average momentum $\langle p\rangle$ decreases very slowly in time in this regime, the original wave packet can travel across the lattice, maintaining its original shape. This feature is the key ingredient for quantum transport and state transfer, where one wants a quantum state to evolve across a complex network, without losing its quantum properties, so that its quantum information content can be recovered from another point in the network.

Therefore, we have again evidence of how the introduction of space correlations in the noise helps preserving dynamical properties better than in the spatially uncorrelated case. This can surely be exploited to design protocols for state transfer and communication across networks.
\par
{\it Conclusions -} We have addressed in detail the effects of spatial
correlations on the dynamics of continuous-time quantum walks on noisy
percolation lattices. Our model, which allows us to address memory effects
and transport properties, is based on a stochastic time-dependent Hamiltonian,
where the hopping amplitudes between adjacent nodes are described as local
random-telegraph processes, which themselves show spatial correlations.
\par
Our results show that classical spatial correlations in the noise
make quantum features of the CTQW more robust. More specifically,
we have provided evidence that the presence of strongly spatially
correlated noise induces robust memory effects on the quantum map,
as compared to the case of uncorrelated RTN. Furthermore, spatial
correlations lead to localization-breaking, i.e. make the walker
able to spread over the network and to reach distant nodes while
still undergoing non-Markovian dynamics. Finally, we have shown
that spatially correlated RTN improves transport properties of an
initially traveling Gaussian packet compared the analogue
uncorrelated case.
\par
Our analysis provides novel insight into the effects of spatially
correlated noise on simple graphs and represents a first step into the
understanding of the role of correlated fluctuations on complex networks,
which, in turn, are extremely relevant to several quantum information and
computation task, such as quantum algorithms, quantum communication and
models for realistic transport across distant nodes.
\par
This work has been supported by EU through the collaborative H2020
project QuProCS (Grant Agreement 641277).

\bibliography{spaqw10.bib}

\onecolumngrid
\appendix
\section{Appendix}
In this appendix we present the MATLAB/Octave code that simulates the dynamics of the quantum walk subject to random telegraph noise, with spatial correlations.

The RTN can be obtained from the Poisson process. If $P$ is the variable describing the latter, then the RTN is simply given by $X = (-1)^P$, i.e. it switches its state at each event of the Poisson process.
For a Poisson process with rate $\gamma$, the time intervals $\delta$ between two events
are independent and exponentially distributed with mean $\gamma^{-1}$. So the probability distribution for $\delta$ is an exponential distribution: $p(\delta) = \gamma \exp(-\gamma\delta)$. The cumulative distribution function is $F(\delta) = 1 - \exp(-\gamma \delta)$.
Thus we can generate the time intervals between events by drawing $\delta$ from the probability distribution above.
By inverting the cumulative function, we obtain that $\delta = -\log(R) / \gamma$, where R is drawn from a uniform distribution in $[0,1]$.

The noise domains are generated according to the prescription presented in the main text. Each domain is labeled by an integer number. The $N \times R$ matrix \lstinline{latticeDef} associates, for each of the $R$ realizations of the noise, each site with its corresponding domain number.

The output of the function is an object containing a vector of time instants \lstinline{t} and the cell array \lstinline{rhoAvg}, containing $\bar\rho$ at each time instant. All the quantities of interest can be evaluated from $\bar\rho$. The code below assumes a particle initially localized in the middle of the lattice. An initial Gaussian wavepacket can also be considered by suitably modifying the initialization of \lstinline{psi}.

\begin{lstlisting}[language=MATLAB,deletekeywords=gamma]
function qw = qw_disorder(varargin)
% QW_DISORDER Simulates a 1-particle 1-d quantum walk with RTN noise and disordered
% domains
%
%   qw = qw_disorder() uses default values for the parameters and returns a
%   QuantumWalk struct (see below)
%
%   qw = qw_disorder('param1',value1,'param2',value2, ...) allows to set custom
%           values to the parameters
%
%   PARAMETERS
%
%   latticeSize         size of the lattice (default 100)
%   noiseRealizations   number of noise histories to average over(def 500)
%   time                rather selfexplanatory  (default 10)
%   gamma               switching rate of the RTN (default 1)
%   p                   the probability of correlation between two sites
%                       (default 0)
%   noiseAmp            Amplitude of the noise wrt the coupling (default 1)
%   onSiteEnergy        selfexplanatory (default 2)
%   coupling            Couplign between first neighbors (default 1)
%   DysonOrder          Order of expansion of the Dyson series of U (def 4)
%   jumpProb            the prob. of a jump in a time step (default 0.02)
%   seed                set the seed of the random number generator
%
%   RETURNS
%
%   A struct containing the above parameters and the fields
%
%   t           time vector
%   rhoAvg      A cell array containing the average density operator at
%               each time instant

    % Argument parsing
    ip = inputParser;
    addParameter(ip,'noiseRealizations',500, @isnumeric);
    addParameter(ip,'latticeSize',100,@isnumeric);
    addParameter(ip,'time',10, @isnumeric);
    addParameter(ip,'jumpProb',0.2,@isnumeric);
    addParameter(ip,'gamma', 1., @isnumeric);
    addParameter(ip,'noiseAmp',.9,@isnumeric);
    addParameter(ip,'onSiteEnergy',2,@isnumeric);
    addParameter(ip,'coupling',1,@isnumeric);
    addParameter(ip,'DysonOrder',8,@isnumeric);
    addParameter(ip,'seed',4,@isnumeric);
    addParameter(ip,'p',.0,@isnumeric);

    parse(ip,varargin{:});

    %% Parameters
    qw.N = ip.Results.latticeSize;  % Lattice size
    qw.p = ip.Results.p;
    qw.noiseRealizations = ip.Results.noiseRealizations;

    qw.time = ip.Results.time; % Total evolution time

    % jumpProb specifies the probability to have a jump in the timestep dt
    % It is used to determine the appropriate dt so that we don't miss jumps of
    % the fluctuators, so it must be low (e.g. 0.2 or less)
    qw.jumpProb = ip.Results.jumpProb;

    % Switching rate
    qw.gamma = ip.Results.gamma;

    % Initial position of the particle
    qw.initialPos = floor(qw.N / 2); % Particle localized in the center

    % Array that specifies the spatial noisy domains
    % Each number represents a noise realization. If two sites have the same
    % number then their noise is correlated

    qw.latticeDef = cumsum([ones(1,qw.noiseRealizations); ...
        (rand(qw.N -1,qw.noiseRealizations) < 1-ip.Results.correlation)]);

    domainIndices = qw.latticeDef;

    % Parameters of the Hamiltonian
    qw.onSiteEnergy = ip.Results.onSiteEnergy;
    qw.coupling = - ip.Results.coupling; % First-neighbor coupling strength
    qw.noiseAmp = ip.Results.noiseAmp * qw.coupling; % Noise amplitude

    Hdiag = qw.onSiteEnergy * ones(qw.N,1);  % On-site energy

    qw.DysonOrder = ip.Results.DysonOrder; % Order of expansion of the Dyson series for
                                           % U = exp (- i H dt)

    % Set the seed of the random number generator
    rng(ip.Results.seed, 'twister')

    % dt for each time step (must be much smaller than the corr. time of
    % the RTN because otherwise we miss jumps)
    dt = min(.5, qw.jumpProb / qw.gamma);
    qw.t = linspace(0, qw.time, floor(qw.time / dt)); % time vector
    dt = qw.t(2) - qw.t(1); % Adjust dt so that we have the exact number of timesteps
    timesteps = length(qw.t); % Number of timesteps

    % Initial state of the system
    psi = zeros(qw.N, qw.noiseRealizations);
    psi(qw.initialPos, :) = 1; % Initially localised particle

    % Function that returns the intervals between the next jumps of the rtn.
    % It draws n x m numbers from an exponential distribution
    rtn_dt = @(n,m) - log(rand(n, m)) / qw.gamma;

    % Count the number of spatial domains
    domainCount = domainIndices(end, :);

    % Initial noise coefficients (equal probability of being +/- c0

    % A matrix of randomly chosen +1 and -1
    pm = 2 * randi(2, max(domainCount), qw.noiseRealizations) - 1;
    nu = qw.noiseAmp * pm(domainIndices); % Initial noise coefficients

    r1 = circshift(1: qw.N, [0,  1]);   % [N, 1, ..., N - 1]
    l1 = circshift(1: qw.N, [0, -1]);   % [2, ..., N, 1]

    % Next jump times
    deltat = rtn_dt(max(domainCount), qw.noiseRealizations);

    %% Output variables
    % Density operator
    [qw.rhoAvg{1:timesteps}] = deal(zeros(qw.N));
    qw.rhoAvg{1} = psi(:,1) * psi(:,1)';

    %% Simulation loop
    for ti = 2 : length(qw.t) % Time loop
        for ni = 1 : qw.noiseRealizations % Noise realizations loop

            % At each time step we update the first band diagonal of the Hamiltonian
            Hi1 = qw.coupling + nu(:,ni);

            % We evaluate the evolved psi by Dyson-expanding U, up to a given
            % order
            kQ = psi(:,ni);
            for k = 1 : qw.DysonOrder
                kQ = (-1i * dt) / k * (Hdiag.*kQ + Hi1.*kQ(r1) + Hi1(l1).*kQ(l1));
                psi(:,ni) = psi(:,ni) + kQ;
            end

            % We check which of the fluctuators have jumped. If they did, we
            % flip their state and calculate the next jump time, updating the
            % jump-time vector
            jumpdomains = qw.t(ti) > deltat(:,ni);
            jds = sum(jumpdomains);
            if jds > 0
                jumps = qw.t(ti) > deltat(domainIndices(:,ni),ni);
                nu(jumps,ni) = - nu(jumps,ni);
                deltat(jumpdomains,ni) = deltat(jumpdomains,ni) + rtn_dt(jds,1);
            end

            % We build the average density operator
            qw.rhoAvg{ti} = qw.rhoAvg{ti} + psi(:,ni) * psi(:,ni)'/qw.noiseRealizations;
        end
    end
end
\end{lstlisting}

\end{document}